# Design, Fabrication, and Measurement of a Hemispherical Multi-Layer Band-Pass Frequency Selective Surface

Ali Tehranian, Jordan Budhu, *Senior Member IEEE*, Casey Perkowski, Lance Sookdeo,
Kenneth H. Church, *Member IEEE* , Garrett Harris, and Carl Pfeiffer, *Member IEEE*

*Abstract*—A hemispherical multilayer wide-band (7-13 GHz) band-pass frequency selective surface (FSS) is reported. A new design technique based on a Goldberg discretization and unit cell scaling technique is introduced to accommodate the curved profile of the FSS. The FSS is additively manufactured by sequentially printing dielectric layers and metallic patterns until 3 patterned silver-ink surfaces are integrated within a 4.5 mm ($\lambda_0/6$ at 10 GHz) thick ABS hemispherical radome. The diameter and the height of the realized hemispherical FSS are around $5\lambda_0$ and $3\lambda_0$ respectively. Measurements demonstrate a roughly 1.7 dB insertion loss in the passband and 15-20 dB rejection in the stopband. Additionally, a new postprocessing technique is used to suppress the effects of edge diffraction in the measured transmission spectrum. The design process, manufacturing technique, and measurement postprocessing represent novel advancements enabling future conformal frequency selective surfaces.

## I. Introduction

FREQUENCY SELECTIVE SURFACES (FSSs) are the wavefront analog of lumped element filters. Numerous examples of FSS structures composed of various element shapes on single or multi-layer platforms have been reported [1-3]. The majority of FSS designs presented in the scientific literature are planar [4-11]. The drawback of planar FSSs is they cannot be utilized in conformal scenarios and they lack angular coverage allowing unfiltered waves to be received through the antennas sidelobes. To circumvent this issue, FSSs with curved profiles, are proposed in [12-18]. The curved-profile FSSs presented in [12] and [13] suggest through simulation that curved structures can provide the desired filtering response along with increased angular coverage. In [14-18], curved FSS prototypes were experimentally investigated. A single FSS layer with a passband of 10 GHz sandwiched between three dielectric layers is presented in [14] with the largest characteristic dimension of the FSS being $10\lambda_0$. The fabrication method in the referenced work is based on the thermal fusing of piecewise planar sections of planar FSS segments. A dual-band composite single layer FSS with a height of 53 cm and with transmission frequencies of 15 and 33 GHz was reported in [15]. Flexible copper film is used to realize the FSS screen which is later bonded to the curved profile of the radome. A wide-band 3-layer FSS with a 1-meter height operating from 7-17 GHz is reported in [16]. More recently, in [17] and [18], methods for projecting initially flat FSS designs onto curved radome platforms were presented. These mapping techniques attempt to correct for distortion introduced by the mapping. In the referenced works, the fabrication was performed manually by hand painting conductive inks onto 3D printed curved substrates. In these works, curved FSSs with wideband or dual-band filtering characteristics have been experimentally demonstrated, however, their fabrication approaches required multi-step and multi-process approaches.

In this paper, the design, fabrication, and experimental verification of a multilayer FSS containing three embedded highly conformal metallic layers manufactured in a single novel process is presented. The hemispherically shaped FSS has a band pass frequency response covering the entirety of X-band. Instead of adopting a fixed unit-cell design and projecting it to the hemispherical platform, the FSS is designed directly onto the curved surface by discretizing the hemisphere into a Goldberg polyhedron. A Goldberg polyhedron is a hexagonal tessellation of the surface of a sphere. Hence, in section II, the design of a three-layer hexagonal FSS unit cell consisting of an inductive grid layer sandwiched between two capacitive patch layers is presented. The design involves a combination of circuit models of FSS layers, transfer matrix methods, and unit cell responses taken from full-wave simulation tools. As the Goldberg polyhedron discretization produces different sized hexagons in the tessellation, a novel scaling approach is used to map the hexagonal unit cell design onto each hexagon. The scaling approach varies the line and gap widths of the inductive and capacitive traces proportional to the hexagon size such that the unit cell sheet impedance is maintained across all hexagons produced in the Goldberg polyhedron tessellation. In this manner, a single unit cell design can be mapped to the entire surface. In section III, full-wave simulations of the CAD model of the completed hemispherical FSS are presented. It is shown that despite the structure's curvature, it attains excellent agreement with the band pass response of the planar unit cell.

In section IV, a planar version of the proposed multilayer bandpass filter manufactured using the same embedded metallic layer approach is introduced for comparative purposes showing the hemispherical FSS attains far superior angular coverage.

Then, in Sections V and VI, the hemispherical FSS is fabricated through additive manufacturing and measured. In this section, a novel postprocessing technique is introduced which can suppress the effects of edge diffraction post measurement. Using this novel technique, excellent agreement between simulated and measured values in both the passband and deep in the stopband is demonstrated. Finally, some concluding remarks are presented in section VII.

## II. Band Pass Unit Cell Design

### A. Goldberg Discretization

There are several methods to tesselate a spherical surface. The Goldberg discretization technique is one example that tessellates the surface of a spherical object with only hexagonal and pentagonal cells [19, 20]. A Goldberg polyhedron is denoted by $GP(m,n)$. Indices $m$ and $n$, denote locations of pentagons on the surface. By starting from one arbitrary pentagonal cell and moving $m$ steps in one direction followed by a 60° rotation and an additional $n$ steps in the new direction, a different pentagonal cell will be reached. For example, a soccer ball is described by a $GP(1,1)$ Goldberg polyhedron. Note, the Goldberg polyhedron discretization always results in exactly twelve pentagonal cells [19, 20].

In the work of this paper, a hemispherical multilayer FSS is designed. The smallest (innermost) diameter of the hemispherical structure is 145 mm, while the middle and the outermost layers have diameters of 147.5 mm and 150 mm respectively. The hemispherical layers were discretized using an open-source tool called *Antiprism* with $GP(20,0)$ [21]. This tool generates output files containing vertex coordinates and connectivity lists of the hexagons forming the Goldberg polyhedron. Figure 1(a) depicts the 150-mm diameter sphere modeled by $GP(20,0)$. Some of the observable pentagonal cells in this view of $GP(20,0)$ are marked with red color.

An important observation obtained from Fig. 1(a) is that the hexagonal cells are not all the same size. The hexagons closer to each of the twelve pentagons have smaller edge lengths, while the ones located farther have larger edge lengths as depicted in Fig. 1(b).

Due to the symmetry inherent to the $GP(20,0)$ there exists an irreducible section of the surface such that the entire polyhedron can be constructed through a series of mirroring and rotation operations. The irreducible section for the outermost layer is shown in Fig. 1(b). A mirror operation across the $yz$-plane generates a 1/5 sector of the polyhedron. Thus, the Boolean addition of the rotation of this 1/5 sector along with its mirror image through the angles of 72°, 144°, 216°, and 288° generates the entire hemispherical polyhedron. Figure 1(b) also indicates the short-diagonal size $p_2$ of each hexagonal cell on the irreducible section. It is found that the range of hexagonal unit cell diameters is 3.22 mm $\leq p_2 \leq$ 4.76 mm, considering all three layers.

### B. Hexagonal Unit Cell Design

Cascading non-resonant capacitive and inductive FSS plates can form a second-order bandpass filter [2, 22, 23]. The most common non-resonant elements for realizing reactive surfaces are capacitive patches and inductor wire-grid [24]. As per the Goldberg discretization, a hexagonal unit cell design is chosen. Therefore, as shown in Fig. 2(a), a hexagonal lattice with long diagonal size of $p$ and short diagonal size of $p_2 = p/\sqrt{3}$ is assumed for the unit cell design.

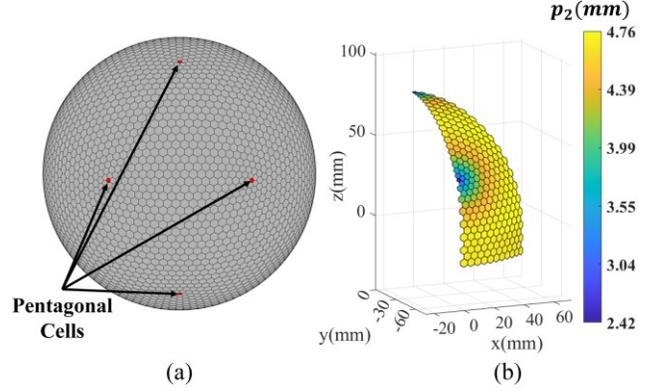

Fig. 1. (a) The $GP(20,0)$ Goldberg polyhedron of diameter 150 mm. The pentagons are shown in red. (b) The irreducible section of the outermost layer. The color bar indicates the size of the short-diagonal of each cell.

A hexagonal wire-grid with the wire-width of $w_L$ following the perimeter of the unit cell as shown in Fig. 2(b) is used for the inductive layer. This layer is then sandwiched between two capacitive layers. For the capacitive layers, hexagonal metallic patches are used [2]. To ease fabrication, the patches are realized as hexagonal wheel-spoke geometries as depicted in Fig. 2(c).

The width of the metallic traces is $w_C$ and the gap between adjacent capacitive elements is $g$. The thin metallic traces are modeled as a lossy metal with conductivity $\sigma = 10^6\ S/m$ in the simulation models to adhere to the conductivity of the silver paste used to fabricate the FSS. For the dielectric spacers, ABS with a relative permittivity of $\varepsilon_r = 2.4$ and loss-tangent of $\tan\delta = 0.006$ is used [25]. The spacers have a fixed thickness of $d_1 = 1.25$ mm. Lastly, protective ABS encapsulation layers with the thickness of $d_2 = 1$ mm are placed on the top and bottom surfaces of the structure as shown in Fig. 2(d).

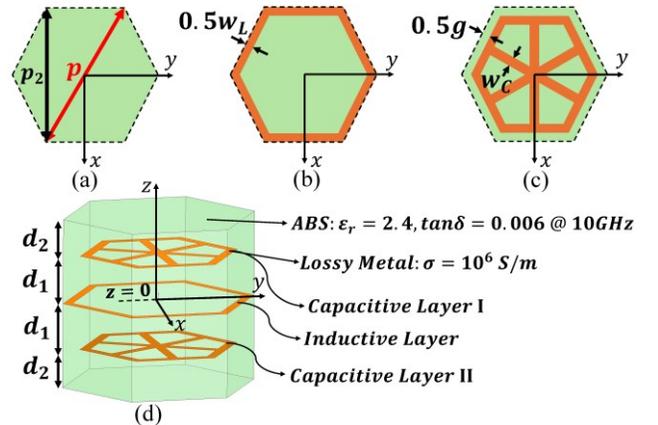

Fig. 2. The proposed multilayer FSS. (a) Unit cell outline, (b) Inductive FSS layer, (c) Capacitive FSS layer, and (d) Perspective view of the multilayer FSS. The dielectric regions made of ABS are shown in green, and the metallic traces are shown in orange.

The full unit cell is shown in Fig. 2(d) and has an overall thickness less than $\lambda_0/6$ at the center for the passband (10 GHz). It is shown in [22] that such a multilayer structure composed of cascaded FSSs and dielectric layers can be easily modeled with shunt purely reactive elements in series with transmission lines representing the dielectric layers. To obtain the frequency response of such a system one can multiply the transfer matrix of each section to calculate the overall transfer matrix at each frequency (see Appendix A). By changing the capacitance and the inductance values of the FSSs, a variety of band pass responses can be achieved (See additional materials in reference [26] for three examples). Accordingly, after performing parameter sweeps over the shunt inductance, $L$, and capacitance, $C$, in the circuit models, the response given by $C = 78$ fF, $L = 1.66$ nH is selected as the design starting point since they provide an acceptable response.

The next step is to map the obtained values for the desired capacitance and inductance, to geometrical parameters of the hexagonal FSSs' unit cells. By appropriately mapping the geometrical features of an inductive or a capacitive square lattice into their hexagonal counterparts, one can use the same mathematical formulas to estimate the respective reactance of the hexagonal FSS layers [1, 2]. The relations for hexagonal wire-grids and metallic patches are given in the additional materials of reference [26]. To account for mutual coupling between the adjacent FSS layers, Ansys HFSS was used to optimize the geometrical parameters of the FSS layers. The final parameters of the FSS layers with $p_2 = 4.5$ mm are reported in Table I. Although there are some discrepancies between the two methods, the simulated unit cell response is sufficiently acceptable. The difference between the two methods stems from the circuit model being only an approximation of the actual unit cell that neglects the material losses in the circuit model, slight frequency dependent behavior of FSS layers, and evanescent mode coupling between the metallic layers. It is also important to investigate the FSS sensitivity to the incidence angle. A similar simulation setup is employed to study this matter. The result of this study and a thorough discussion about it are presented in additional materials of [26]. It is shown that our designed FSS can provide a robust response for large incidence angles as high as 45°.

Parameter sweeps were performed in HFSS with the obtained design parameters reported in Table I as the starting point to find scaling laws for the geometrical parameters of FSS layers such that the sheet impedance of each layer remains constant as the short hexagonal diameter $p_2$ is varied.

TABLE I
GEOMETRICAL PARAMETERS OF THE CASCADED FSS UNIT CELL

| PARAMETER | VALUE |
|---|---|
| $p$ | 5.19 mm |
| $p_2$ | 4.5 mm |
| $w_L$ | 0.22 mm |
| $g$ | 0.8 mm |
| $w_C$ | 0.25 mm |

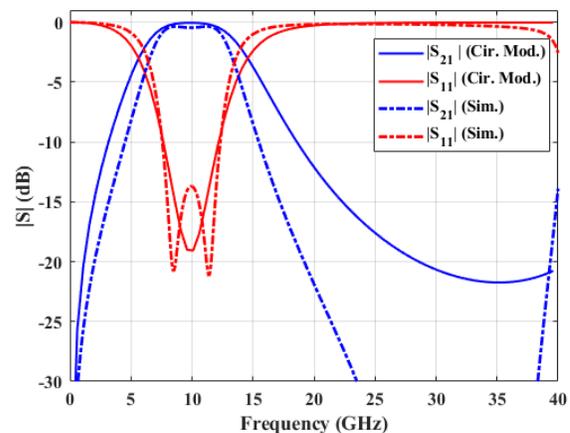

Fig. 3. Comparison of the S-parameters of the FSS equivalent circuit model and the full-wave simulated FSS structure. The circuit model curves are plotted with solid lines and the simulated results with dotted-dashed lines. Red curves depict $|S_{11}|$, and blue curves depict $|S_{21}|$.

Figure 3 shows the S-parameters of the simulated unit cell along with the filter response obtained from the circuit model. Recall, this is necessary as the Goldberg discretization scheme of Fig. 1(b) produces a range of hexagon sizes. Hence, these scaling laws enable the unit cell response depicted in Fig. 3 be maintained across all the hexagons in the tessellation. By varying the gap-width of the capacitive layer and the width of metallic traces for the inductive layer, $g$ and $w_L$ respectively, the equi-impedance sheet scaling laws were empirically found to be:

$$g(p_2) = \frac{0.3}{0.8} p_2 - 0.88 \text{ mm} \quad (1)$$

$$w_L(p_2) = \frac{0.2}{0.85} p_2 - 0.6 \text{ mm} \quad (2)$$

By plugging in the hexagon diameters in Fig. 1(b) into (1) and (2), we obtain the metallic layer geometries of each layer of the multilayered FSS.

Finally, to form the final structure, $GP(20,0)$ is bisected and joined to a cylinder of the same diameter and 25 mm height. Then the Goldberg unit cell pattern along the equator of the hemisphere is extended/continued down the cylindrical surface to obtain the final discretized hemispherical FSS (see Fig. 6 for preview).

C. *Pentagonal Cell Design*

As mentioned previously, the Goldberg discretization scheme also produces exactly 12 pentagonal cells. Since the sphere is cut in half to form the final FSS geometry, the structure has only six isolated and identical pentagonal cells across the surface. A similar unit cell structure is adopted for the pentagonal unit cells (see Fig. 4).

The line width of the pentagonal cells on every layer is fixed at $w_p = 0.25$ mm and the gap width for the capacitive layer is fixed at $g_p = 0.38$ mm for the outermost and the innermost capacitive layers. The pentagonal side length is 1.51 mm for the inductive layer and the diagonal size of pentagonal capacitive cell is 2.22 mm and 2.13 mm for the outermost and innermost layers, respectively.

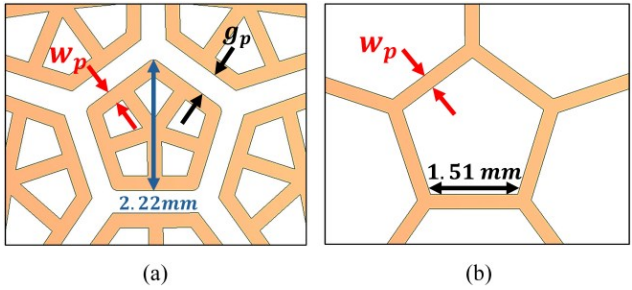

Fig. 4. Pentagonal cell geometry. (a) Outermost capacitive layer geometry. (b) Inductive layer geometry.

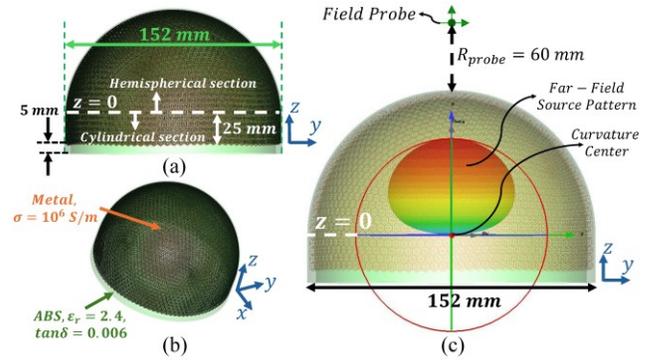

Fig. 6. Final multilayer hemispherical FSS structure, (a) Side view, and (b) Perspective view. The dielectric region is shown in semi-transparent green color, and the metallic regions are shown in orange. (c) The full-wave simulation configuration set up in the IE-solver environment of CST MWS. A far-field source is placed at the center of the spherical part of the structure at $z = 0$ plane, and an Electric-field probe is placed in the structure boresight direction at a distance of $R_{probe} = 60\ mm$ from the top of the structure.

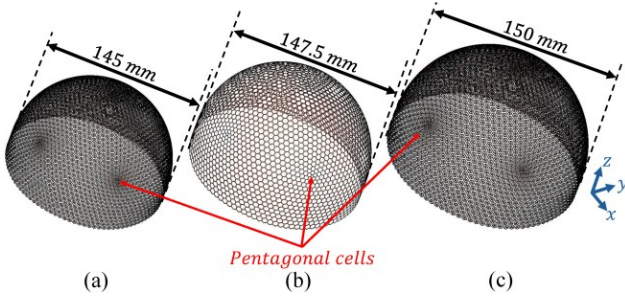

Fig. 5. Perspective view of individual hemispherical FSS layers. (a) Innermost capacitive layer, (b) Inductive layer in the middle, and (c) Outermost capacitive layer. Some of the pentagonal cells are denoted in these figures with red arrows.

Both the capacitive and inductive layer dimensions were chosen to maintain the same sheet impedance as the hexagonal unit cells.

### D. Final FSS CAD Model

Figures 5(a)-5(c) respectively show the individual models of innermost, middle and outermost FSS layers drawn in the CST MWS CAD environment. In Figs. 6(a) and 6(b), the full CAD model is shown in side and perspective views respectively, including the dielectric regions. The height of the dielectric layers at the end of the cylindrical section is extended to 5 mm, to facilitate the installation of mounts or holders for experimentation. The dielectric regions are illustrated in a semitransparent manner in this figure. With this CAD model, full-wave simulations of the full hemispherical FSS can be performed and are described in the next section.

### III. FULL-WAVE SIMULATION OF THE STRUCTURE

The simulated performance of the hemispherically shaped multilayer FSS is obtained using the integral equation (IE) solver in CST MWS. Figure 6(c) depicts the simulation configuration. The FSS is excited by a far-field source placed at the center of curvature of the hemisphere at $z = 0$ plane, the interface plane where the hemispherical section meets the cylindrical section.

The Raised Cosine feed model is used in our simulations as the far-field source. The raised cosine feed model can be used to approximate the radiation pattern of an actual feed antenna. The symmetrical raised cosine feed is described by [27]

$$\vec{E}^{inc} = E_0 \frac{e^{-jkr}}{r} \cos^q \theta\, (\hat{\theta} \cos\phi - \hat{\phi} \sin\phi) \quad (3)$$

where $k = 2\pi f/c_0$ is the free space wavenumber at the simulation frequency $f$, $c_0$ is the speed of light in vacuum, and $q$ is the parameter controlling the directivity of the feed. The value of $q$ is determined so that the directivity of the feed model and its 3-dB beamwidth are matched to those of RF Spin QRH50 quad-ridged horn antennas used in the measurements. The details of the procedure for the determination of parameter $q$ is explained in the additional material of [26].

Next, a field-probe is placed above the FSS at a distance of $R_{probe} = 60$ mm along its boresight direction ($z$-axis). This probe records the absolute electric field value at its location. By reading the probe's data at different frequencies and normalizing to the probes reading when the FSS is absent, one can obtain the normalized transmission spectrum of the FSS.

### A. Boresight Normalized Transmission Spectrum

The normalized transmission spectrum is calculated and plotted in Fig. 7 superimposed over the response of the unit cell. The performance of the hemispherical FSS is in excellent agreement with the frequency response of the unit cell. Not only does the hemispherical FSS possess the desired band pass feature in the targeted frequency range but also, its passband transmission efficiency and its roll-off slope agrees with the unit cell response.

To gain insight into the performance of the hemispherical FSS, its polar gain patterns at 8, 12, 16 and 25 GHz in $\phi = 90°$ cut are calculated and presented in Fig. 8 along with the gain patterns of the Raised Cosine model at 8 and 25 GHz. The figure shows that at the frequencies of 8 and 12 GHz, which are at the passband edges, the FSS passes the waves emitted from the source. Comparing the magnitude of the forward and back lobes at 8 and 12 GHz suggests a better transmission at 12 GHz which agrees with the obtained normalized transmission spectrum shown in Fig. 7. By increasing the frequency beyond 12 GHz, one enters the stopband of the filter. This is also evident by looking at the gain patterns at 16 and 25 GHz where the back lobes are stronger compared to the boresight lobes. At 16 GHz which is still close to the bandpass region, the magnitude of the back lobe is only slightly larger than the boresight lobe; but at 25 GHz resided in deep band-stop region, the structure acts as a mirror and reflects most of the power towards the rear direction ($\theta = 180°$).

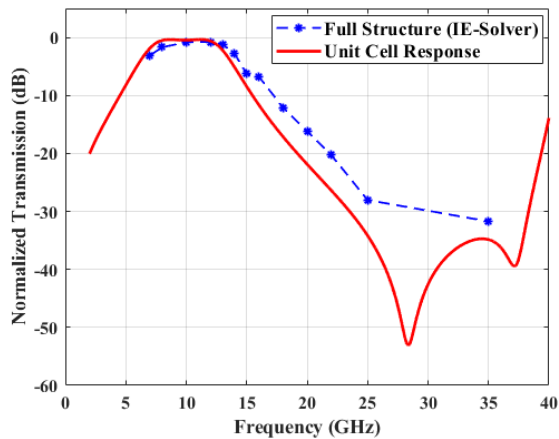

Fig. 7. Comparison of the normalized transmission of the hemispherical FSS with the obtained unit cell response.

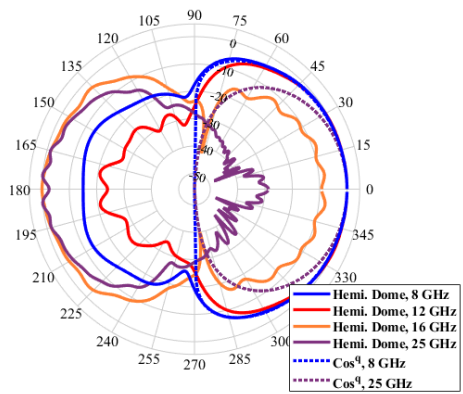

Fig. 8. The obtained polar gain patterns of the structure in $x = 0$ plane ($\phi = 90°$) plane, at 8, 12, 16 and 25 GHz. It is evident that outside of the passband region, the structure reflects the waves towards the rear direction.

## B. Angular Coverage and Scanning Performance

As previously mentioned, a hemispherical FSS has superior angular coverage. Another advantage of the spherical FSS is that the radiation pattern of a transmitting antenna placed inside the FSS can be scanned while maintaining filtering. Here, the filtering efficiency of the structure versus different scan angles of the feed and the probe locations are assessed. In section IV, we will compare the angular coverage and scanning performance to the more traditional planar FSS approach.

The simulation configuration is illustrated in Fig. 9. As shown in the figure, in addition to the probe in the boresight direction of the structure ($\theta_{probe} = 0°$), several probes are placed at different angles relative to the boresight direction at $\theta_{probe} = 30°, 45°, 60°$ and $90°$, but with the same distance from the far-field source and the FSS surface. The probes distances from the far-field source are $R = 136$ mm. By rotating the feed around the x-axis with $\theta_{Feed}$ angle relative to z-axis, its main lobe also rotates with the same angle. Following the same approach as was done for the boresight angle case, by reading the probes recorded data once with the FSS in place and once without it, one can obtain the absolute field values and hence the normalized transmission for different angles of detection ($\theta_{probe}$) and different angles of excitation ($\theta_{Feed}$).

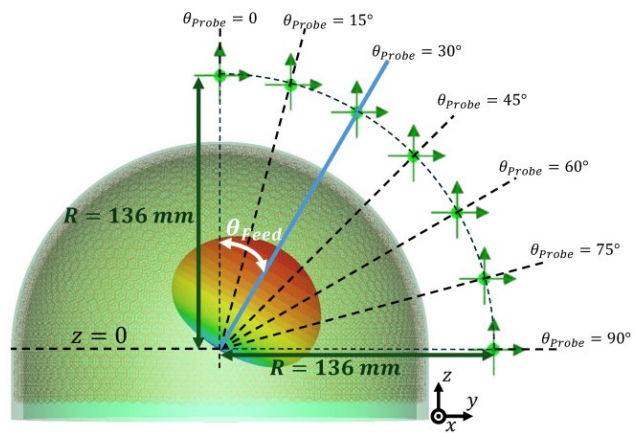

Fig. 9. The simulation configuration for the study of the effect of the angle in the hemispherical FSS performance. In different simulation scenarios, the far-field source is rotated around the x-axis with the angle $\theta_{Feed}$. Several Electric-field probes were placed around the structure, all with a same distance to the excitation source ($R = 136$ mm) surface but at different angles ($\theta_{probe}$).

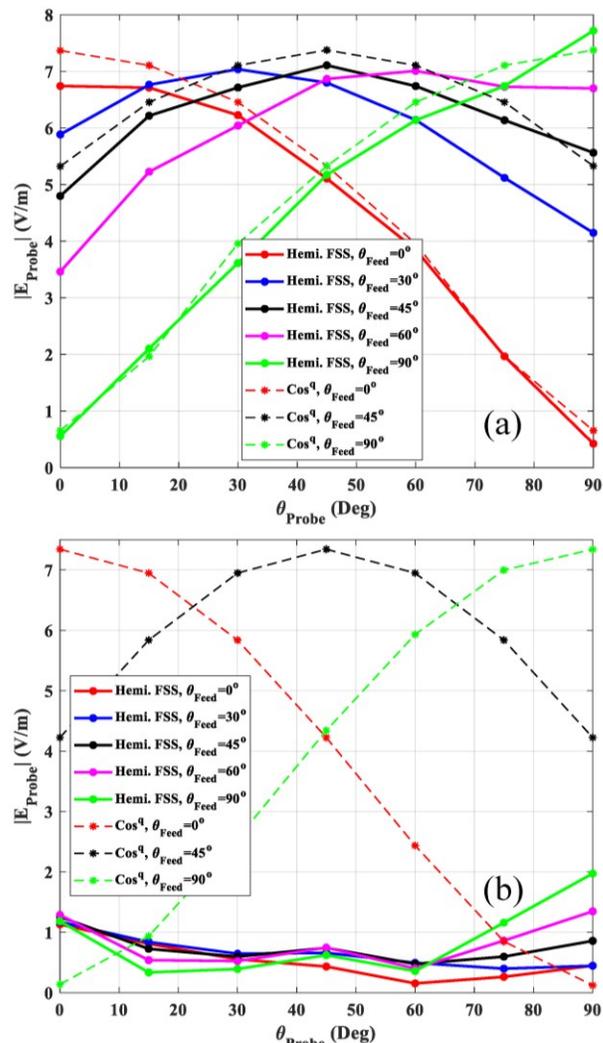

Fig. 10. Absolute Electric-field values at the probe locations around the hemispherical FSS, i.e. same distance from the excitation source but at different angles ($\theta_{probe}$) at (a) 10 GHz, and (b) 20 GHz.

However, since very low transmission coefficient is expected when the feed angle is far from the probe angle, especially for the frequency samples in the stopband range, the absolute electric field values (not-normalized) are presented and compared here.

The absolute electric field values at the probe's locations are calculated for $\theta_{Feed} = 0, 30°, 45°, 60°$ and $90°$ at 10 GHz and 20 GHz and are shown in Figs. 10(a) and 10(b), respectively, along with those of the Raised Cosine source alone directed towards $\theta_{Feed} = 0, 45°, 90°$ angles. Figure 10(a) shows that the feed pattern is maintained as the feed pattern is scanned within the hemispherical FSS. Another interesting point is that even though the maximum field value is almost the same for different feed angles, its value at the completely horizontal feed direction ($\theta_{Feed} = 90°$) is slightly higher. This phenomenon stems from the scattering effects around the structure edge and the fact that in this case a part of the feed power can reach the probe location ($\theta_{probe} = 90°$) without passing through the FSS.

## IV. A Planar FSS and its Comparison to the Hemispherical FSS

The performance of a planar version of the band pass FSS structure is compared to the hemispherical version in this section. A perspective view of the planar FSS band pass filter as well as the 3D simulation configuration setup in the IE solver of CST MWS is presented in Figs. 11(a) and 11(b) respectively. The unit cells composing the structure are the same as shown in Fig. 2(a)-2(d). The outline shape of the planar FSS itself is a large hexagon.

The long diagonal length of the FSS along the $x$-axis is 180 mm, and its short diagonal length along the $y$-axis is 156 mm.

The far-field source described in (3) is placed under the structure in the center of the simulation configuration and excites the structure from below. The distance between the far-field source and the planar FSS is 76 mm. The normalized transmission is defined as the difference of the recorded field value at the probe location with the structure in place and without it in dB scale. The normalized transmission of the planar structure along with that of the hemispherical FSS, both under normal incidence, and the unit cell response are calculated and plotted in Fig. 12. As expected, the planar structure also performs well in realizing the desired band pass filter.

It is obvious that compared to the hemispherical FSS covering an entire half space above the feeding source ($0 \leq \theta \leq 90°$), the planar FSS covers only a part of this angular space. To analyze the planar FSSs' angular coverage, the effect of feed angle ($\theta_{Feed}$) on the planar FSS response at 20 GHz is assessed. Similar to the angle effect study performed for the hemispherical FSS, several electric field probes are placed around the structure in the $xz$-plane with a same distance of 136 mm to the far-field study. The absolute electric field values at different probe locations with the same distance to the source but different angles relative to structure central axis, $\theta_{probe}$ (see Fig. 11(b)), is obtained for different feed angles and is plotted in Fig. 13.

It is evident that for normal illumination ($\theta_{Feed} = 0°$), the boresight probe shows that the planar FSS is filtering the incoming wave as expected since 20 GHz is in the bandstop region. However, by changing the detection angle towards the edge of the planar structure, i.e. increasing $\theta_{probe}$, the field value increases as the result of unfiltered fields reaching the probes not covered by the planar FSS. This gets worse by increasing the feed angle. In fact, for feed angles higher than 45° and detection angles higher than 60°, there is almost no filtering effect present. For smaller feed and detection angles, the filtering functionality is impaired.

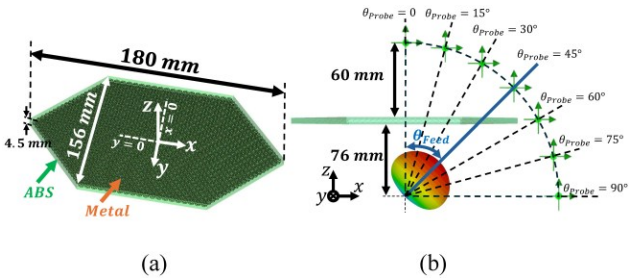

(a)    (b)

Fig. 11. (a) A perspective view of the planar multilayer FSS band pass filter realized by the proposed FSS unit cell. (b) The simulation configuration for the planar structure. In addition to a boresight direction probe located 60 mm above the structure, other probes with the same distance to the far-field source (136 mm) are placed around the structure in $y = 0$ plane but with different angles ($\theta_{probe}$).

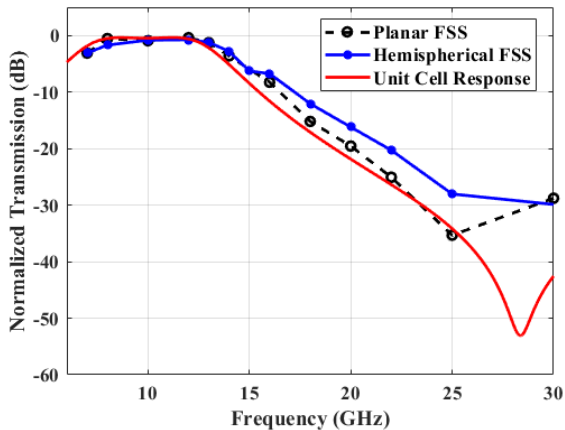

Fig. 12. Comparison between the boresight normalized transmission of the planar structure, the hemispherical FSS, and the unit cell frequency response.

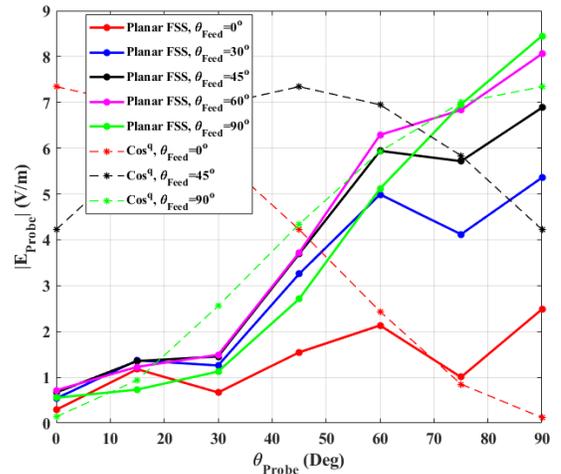

Fig. 13. Absolute Electric-field values at the probe locations around the planar structure, i.e. same distance from the excitation source but at different angles ($\theta_{probe}$) at 20 GHz.

## V. Additive Manufacturing of the Conformal Prototype

Although complex conformal RF designs have been widely created and simulated, manufacturing these designs presents a large challenge due to the complexity and tight tolerances required. As discussed in the introduction, many methods have been shown for producing conformal designs, however these methods are either limited in the curvatures they can work with, the manual processing needed or the number of layers the device can contain. The method presented here provides an additive manufacturing process that can produce multi-material, multi-layer, highly conformal, encapsulated designs for RF devices in an automated fashion and on a single system.

The machine used for manufacturing the hemispherically shaped multilayer FSS was an nScrypt 3Dx-700 system outfitted with a trunnion-style dual-axis rotary configuration. The 5-axis motion-capable machine achieved X, Y, and Z translation using a high-precision gantry, and $\theta$ and $\varphi$ rotations using the stationary dual rotaries. This configuration was specifically chosen to simplify the machine motion required to traverse the hemispherical surface and help increase control of the critical printing gap between the print head and the part.

For conductive layers this gap is 30 microns and can have a large effect on the flow characteristics of the material at the tip and the resulting line width. The system is equipped with multiple tool heads which include the Smartpump™ for microdispensing of conductive layers, the nFD™ for Fused Filament Fabrication (FFF) of dielectric layers and the nMill™ which is used for precision surface machining and smoothing. For all dielectric layers, 3DxTech Black ABS filament is used with a 300 μm internal diameter (ID) pen tip. For all conductive layers a dispensing specific silver paste, ACI materials FS0142, was used with a 125 μm ID pen tip. All milling processes used a 1/8" diameter ball end mill bit.

Initial steps included the translation of design files into manufacturing files. This is typically trivial for planar designs due to existing software tools. For conformal designs, generating tool paths becomes more complicated by the intricacies of traversing 3D surfaces and the need to determine both the position and normal required for the machine along the desired paths. For conductive layers, special versions of the CAD files were created with the element geometry offset by half of the single pass line width of the print head on each side. This ensures the resulting geometry produced during printing matches the required design dimensions.

These offset versions of the CAD files were opened in Rhinoceros 3D (Rhino), a Non-Uniform Rational B-Splines (NURBS)-based CAD software, for processing. An nScrypt-specific Rhino plug-in, which traversed the conformal surface and segmented the tool paths based on deviation from the base surface, was used to generate the manufacturing file for the system. For dielectric layers and milling, a custom program was developed in Grasshopper, a visual programming language integrated with Rhino, to generate spiral paths with equal spacing that covered the hemispherical surface.

These paths were then output as manufacturing files for the system using the same plug-in. From these manufacturing files, the nScrypt system calculated the inverse kinematics required for all paths and added the necessary travel movements, valve commands, and other process controls. For the hemispherically shaped multilayer FSS, the initial portion of the inner encapsulation layer was printed on a separate FFF machine and then mounted onto the 3Dx-700 system. An initial milling pass was performed to smooth the surface, and the inner encapsulation layer was completed by conformally FFF-printing the remaining thickness.

To improve adhesion and minimize deformation across all conformal FFF layers, directed heating was applied using temperature-controlled hot air flow. The finished internal encapsulation layer was subsequently milled, removing the typical bumpy texture of FFF-printed surfaces. This both increases the RF performance [28] and provides a smooth, consistent surface for dispensing. Once the surface is prepared, the first conductive pattern is realized by an approx. 2.5-hour continuous micro-dispensing process. Next the first spacer is conformally printed with FFF which protects and fuses the conductive pattern into the hemisphere's structure.

The first spacer is milled in preparation for the second conductive pattern. This method of microdispensing the conductive pattern, protecting with the dielectric and then milling the surface is repeated to add the second conductive pattern, second spacer, third conductive pattern, and final encapsulation layer, in that order. In total, manufacturing machine time totals roughly 66 hours. This represents a non-optimized process time which can be reduced by an estimated order of magnitude through the use of laser trimming for conductive patterns, larger nozzle diameter and increased extrusion rates for dielectric layers, increased tooling size for milling, and increased print and travel movement speeds. Figure 14 depicts the pictures of the manufacturing process.

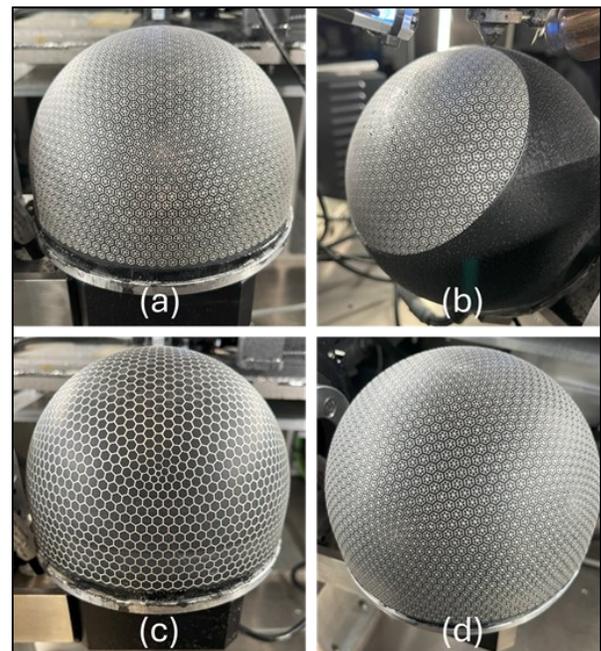

Fig. 14. Manufacturing process pictures: (a) Completed conductive pattern 1, (b) Printing of dielectric spacer 1 showing the encapsulation of the conductive pattern 1, (c) Completed conductive pattern 2, (d) Completed conductive pattern 3.

## VI. Measurement Results

The transmission coefficient of an FSS is traditionally characterized by using lens-antennas on either side that generate a quasi-Gaussian beam illumination to minimize the impact of diffraction. However, sourcing a Gaussian-beam horn with the proper bandwidth, focal length, and spot size is challenging. Therefore, we utilize a simpler setup as shown in Fig. 15. A quad ridge horn (RF Spin QRH50) is integrated in the middle of the FSS and the radiated far-fields measured using an identical horn at the other end of the anechoic chamber.

The measured transmission coefficient is normalized by a calibration measurement with the FSS removed from the measurement setup. In addition to the FSS, the planar FSS is also characterized by mounting it 75 mm in front of the quad-ridge horn antenna. All transmission measurements employ time-gating with a 0.5 ns window to reduce the impact from reflections on the chamber walls. The measured co-polarized transmission coefficient for the planar FSS and hemispherical FSS are plotted in Fig. 16. Also shown in the figure are the simulated results for the unit cell and the planar FSS excited by the Cosine feed model placed 101 mm (4-inch) below the structure. There is general agreement between all three data sets in the sense that there is a passband around 10 GHz and a stopband elsewhere. However, the measurements include additional ripple that we attribute to diffraction around the edges of the sample. This is especially true for the planar FSS measurement where diffraction is expected to be more significant. This reasoning is also corroborated in the simulation data. For reference, the transmission is -10 dB at 10 GHz when the planar FSS is replaced with a similar sized metal plate.

A new measurement technique was recently introduced to combat diffraction effects and improve transmission measurement accuracy of a sample-under-test [29]. Conceptually, the horn antenna at the far end of the chamber is replaced with a Gaussian beam source to reduce the impact of diffraction. However, rather than implementing a physical Gaussian lens-antenna, a Gaussian beam is generated synthetically by post processing far-field measurements at different angles. To begin, the desired Gaussian excitation, $g(x,y)$, is defined at the location of the sample-under-test,

$$g(x,y) = e^{-(\frac{x^2+y^2}{w_0^2})} \quad (4)$$

where $w_0$ is the beam waist radius. For each frequency of interest, a discretized uniformly spaced $\lambda/8$ grid is defined to approximate the continuous desired wavefront. The beam waist radius is set to 69 mm so that 99% of the power of the excitation is incident on the 150 mm diameter sample. The desired excitation ($g(x,y)$) is back-projected into the far-field, $FF_G(\theta,\phi)$, by summing the contributions of each individual point in the excitation,

$$FF_G(\theta,\phi) = \sum_x \sum_y g(x,y) e^{jk\sin(\theta)(x\cos(\phi)+y\sin(\phi))} \quad (5)$$

where $k$ is the free-space wavenumber and we assume an $e^{j\omega t}$ time dependence. Proportionality constants are neglected for the time being because they will eventually be calibrated out in the measurements. This desired far-field excitation is then synthetically excited by weighing the measured far-field of the sample-under-test, $S_{21}^{\text{Measured}}(\theta,\phi)$,

$$S_{21}^{\text{Gaussian}} = \sum_\theta \sum_\phi S_{21}^{\text{Measured}}(\theta,\phi) FF_G(\theta,\phi) \sin(\theta) \quad (6)$$

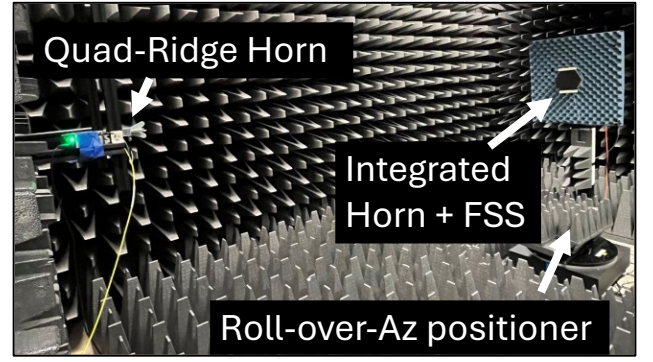

Fig. 15. Measurement setup for characterizing the planar FSS and hemispherical FSS.

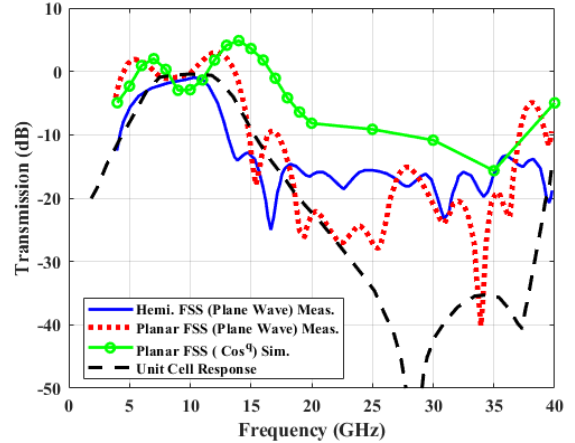

Fig. 16. Measured transmission of the hemispherical FSS and planar FSS when illuminated with a plane wave at normal incidence (along the z-axis). The simulated unit cell response is also included for reference.

The $\sin(\theta)$ term ensures there is a uniform sampling over solid angles of the back-projected field. Eq. (6) represents the effective transmission coefficient from the horn integrated with the sample-under-test, through the sample, and to a Gaussian beam at the other side.

The measured $S_{21}^{\text{Gaussian}}$ of the planar FSS and hemispherical FSS are shown in Fig. 17 plotted over the simulated unit cell response. As expected, the planar FSS transmission closely agrees with the unit cell simulation since the planar FSS is fabricated most accurately. The hemispherical FSS also shows reasonable agreement with the simulated unit cell. Measurements have roughly 1 dB additional loss in the passband at 10 GHz. The measured stop-band insertion loss is around 15-20 dB above 10 GHz, which is lower than simulation. The discrepancy between measured and simulated stop-band insertion loss is attributed to imperfect fabrication of the metal patterns. The new measurement technique was also applied to the simulation data from the planar FSS to corroborate the measured result after postprocessing. As can be seen, the diffraction ripples observed in the transmission spectra as well and the depth of the stopband are improved after the postprocessing.

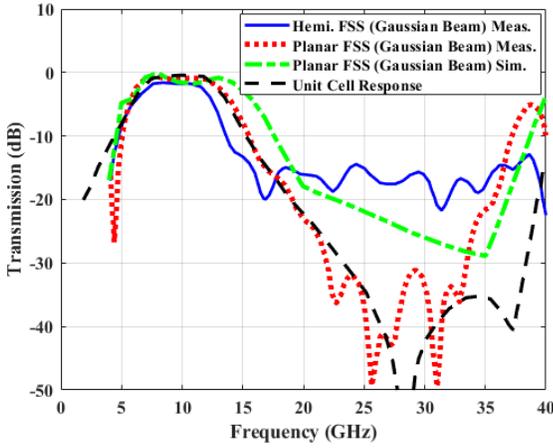

Fig. 17. Measured transmission of the hemispherical FSS and planar FSS when far field measurements are post-processed to synthetically illuminate the sample with a Gaussian beam at normal incidence (along the z-axis). The simulated unit cell response is also included for reference.

These results also demonstrate that the planar FSS diffracts strongly whereas the hemispherical FSS does not. While there is room for improvement, the results demonstrate the proposed approach can realize functional multi-layer FSSs with engineered pass-band characteristics on significantly contoured surfaces.

## VII. Conclusion

A hemispherical multilayer FSS operating as a wideband filter passing the entirety of the X-band is implemented onto a hemispherical platform. A simple and straightforward design approach is offered which can be used for the design and study of multilayered FSS filters. An approach for the discretization of the surface of a sphere into hexagonal unit cells and the subsequent mapping of a hexagonal unit cell onto the hemispherical surface is presented. This mapping scheme maintains the unit cells performance while being scaled to adhere to the Goldberg polyhedrons varying hexagonal dimensions inherent to its tessellation. The hemispherical FSS performance is fully assessed using 3D full-wave simulations. It is shown that the design works excellently in filtering the frequencies in the stopband over a wide range of feed scan angles. The performance of the hemispherical FSS is then compared to a planar version of the same filter. It is shown that the hemispherical FSS outperforms the planar FSS for large incidence angles while it compares well for smaller incidence angles. The fabrication technique for the FSS is also presented. A new measurement technique was proposed to effectively emulate a Gaussian beam excitation. The measured results agree with those predicted by the simulations.

## Appendix A
### Theoretical Synthesis and analysis of Cascaded FSS structures Based on Circuit Model and Transfer Matrix

The equivalent circuit model of the structure is shown in Fig. 18(a). The capacitive FSS layers are modeled with the shunt capacitors ($C_1 = C_2 = C$), the inductive FSS layer in the middle of the structure is modeled with a shunt inductance ($L$) and each of the dielectric layers are modeled with transmission lines denoted by their respective length and characteristic impedance. The free space wave impedance, $Z_0 = 120\pi\ \Omega$, terminates both ends.

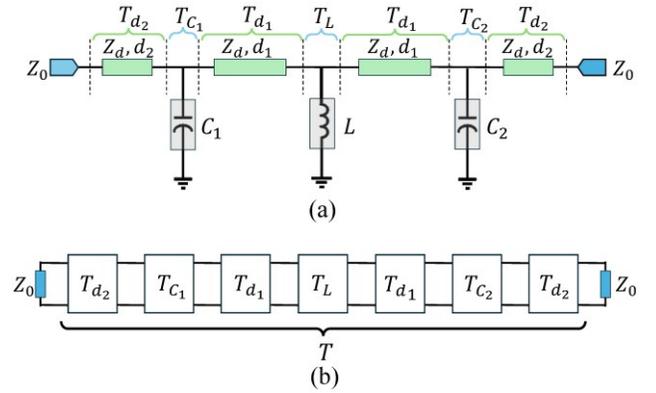

Fig. 18. Equivalent model of the proposed FSS. (a) Equivalent circuit model, and (b) equivalent two-port network model. Each subsection of the structure is separated from its neighbors by dashed lines in the circuit model and with the blocks in the network model.

In this method, each section of the structure and essentially the whole system is treated like a two-port network. The two-port network equivalent model of the proposed unit cell is shown in Fig. 18(b) where each section is replaced by its own matrix block cascaded with other blocks. At both ends, $Z_0$ represents the air as a load for the system. The transfer matrix of a shunt admittance $Y$ as a two-port network can be written as [30]:

$$T_Y = \begin{pmatrix} 1 & 0 \\ Y & 1 \end{pmatrix} \tag{A1}$$

Where for the capacitive layers $Y = j\omega C$, and for the inductive layer $Y = 1/j\omega L$, with $\omega$, $C$ and $L$ being the angular frequency, capacitance of the capacitive layers and the inductance of the inductive layer respectively. For each of the dielectric regions with the thickness of $d_i$ with $i = 1,2$, the transfer matrix can be represented as the following [30]:

$$T_{d_i} = \begin{pmatrix} \cos\beta_d d_i & jZ_d \sin\beta_d d_i \\ jY_d \sin\beta_d d_i & \cos\beta_d d_i \end{pmatrix} \tag{A2}$$

where $\beta$, $Z_d$ and $Y_d$ are propagation constant, characteristic impedance and characteristic admittance of the dielectric regions, and $\beta = \frac{\omega}{c}\sqrt{\varepsilon_r}$ and $Z_d = Z_0/\sqrt{\varepsilon_r}$ where $c$ is the light speed in free space, $Z_0 = 120\pi\Omega$ the free space characteristic impedance, $Y_d = \frac{1}{Z_d}$, and $\varepsilon_r = 2.4$ the relative permittivity of the dielectric material.

The transfer matrix of the whole unit cell can be easily calculated by matrix multiplication of the cascaded blocks. According to the two-port network model of Fig. 18(b), one can write $T$, the overall $2 \times 2$ transfer matrix of the system as:

$$T = \begin{pmatrix} A & B \\ C & D \end{pmatrix} = T_{d_2} \times T_{C_1} \times T_{d_1} \times T_L \times T_{d_1} \times T_{C_2} \times T_{d_2} \tag{A3}$$

After calculating the elements of $T$ matrix at each frequency, the S-parameter matrix of the system at each frequency can be easily obtained. One must note that $S_{11} = S_{22}$, and $S_{12} = S_{21}$, since the proposed structure is symmetric. Now, the S-parameters of the whole unit cell at each frequency can be easily obtained as the following [30]:

$$S_{11} = \frac{A + \frac{B}{Z_d} - CZ_d - D}{A + \frac{B}{Z_d} + CZ_d + D} \quad (A4)$$

$$S_{21} = \frac{2}{A + \frac{B}{Z_d} + CZ_d + D} \quad (A5)$$

It is worth mentioning that since the structure is symmetric and reciprocal, the relation $AD - BC = 1$ is satisfied.

## APPENDIX B

### EXAMPLES OF BAND PASS FILTERS REALIZED BY CASCADED CAPACITIVE-INDUCTIVE-CAPACITIVE FSS STRUCTURE

Here we present some examples of different band pass filter responses realized by the proposed FSS unit cell. The unit cell illustrated in Fig. 2 is composed of two wheel-spoke hexagonal capacitive FSS layers with an inductive wire-grid FSS layer in between, while all these layers are buried inside a dielectric material made of ABS. The transfer function and the procedure for obtaining the frequency response of the FSS filter is thoroughly explained Appendix A. Three Examples are presented here for different values of $L$ and $C$. In all cases, the dielectric properties and their thickness are fixed to the mentioned values in section II, and only different values of $C$ and $L$ are considered. Figure 19 shows the S-parameters of the examples.

The first example marked by solid lines show the filter response of a case with $C = 45\ fF$ and $L = 1.66\ nH$. The second and the third cases are the obtained filters with $C = 78\ fF, L = 1.66\ nH$ and $C = 78\ fF, L = 10\ nH$, which are marked with dashed-lines and dotted-dashed lines respectively. These examples are cleverly chosen close to the final optimized solution suitable for our purpose. As expected, by choosing different values of $C$ and $L$ one can achieve various filter responses. Additionally, it is obvious that specifically for the current work where no specific limitations are assumed for the filter response, there exist numerous acceptable solutions.

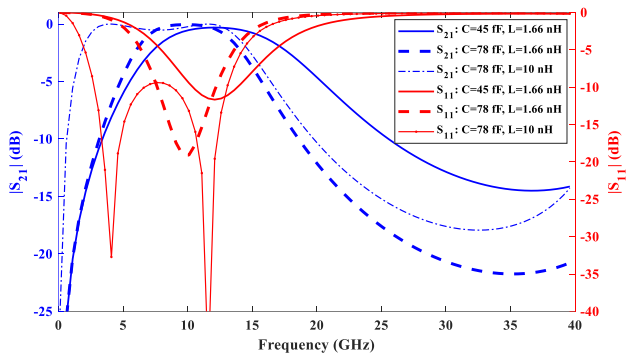

Fig. 19. The S-parameters of three example FSS unit cells with the proposed geometry. Example one has $C = 45\ fF, L = 1.66\ nH$ and is marked with solid line. Example two denoted by dashed line has $C = 78\ fF, L = 1.66\ nH$, and the third example marked with dotted-dashed line is $C = 78\ fF, 10\ nH$. The axis on the left indicates $|S_{21}|$ for the blue curves and the right axis indicates $|S_{11}|$ for the red curves.

## APPENDIX C

### ANALYTICAL FORMULAS FOR HEXAGONAL CAPACITIVE AND INDUCTIVE FSS STRUCTURES

In this section, analytical relations for the capacitance of hexagonal patches and the inductance of hexagonal wire-grids are presented. One can employ these relationships to relate the geometrical parameters of these structures to their respective reactance values. Obviously, these relations are approximate models and require further modifications. However, they are useful for obtaining a starting point for further studies.

Let us assume that we have a hexagonal lattice of patches with hexagonal geometry. All the hexagonal patches are equilateral and identical with the edge length of $R$. The periodicity of the lattice is $D_x$ and $D_y$ in the x-axis and y-axis directions respectively, and the gap width between the edges of adjacent patches is $g = D_y - 2R.\cos 60°$ as shown in Fig. 20(a). The effective capacitance of this FSS can be obtained according to [2, 31]:

$$C = \varepsilon_0 \varepsilon_{eff} \left(\frac{2D_y}{\pi}\right) \ln\left(\frac{1}{\sin\left(\frac{\pi g}{2D_y}\right)}\right) \quad (A6)$$

Where $\varepsilon_0$ is the permittivity of vacuum and $\varepsilon_{eff}$ is the effective permittivity of the medium. The counterpart of this FSS is a metallic wire-grid in a hexagonal lattice with $r$ as the edge length of hexagonal apertures, $w_L = D_y - 2r.\cos 60°$ the width of metallic traces, $D_x$ as the periodicity of the lattice in x-axis direction and $D_y$ the periodicity of the FSS along the y-axis as shown in Fig. 20(b). The effective inductance of such FSS can be easily obtained using the following relation [2, 31]:

$$L = \mu_0 \mu_{eff} \left(\frac{D_y}{2\pi}\right) \ln\left(\frac{1}{\sin\left(\frac{\pi w_L}{2D_y}\right)}\right) \quad (A7)$$

Where $\mu_0$ is the permeability of vacuum and $\mu_{eff}$ is the effective permeability of the medium.

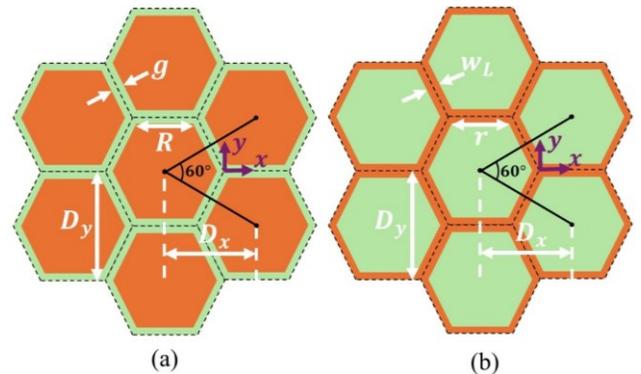

Fig. 20. A hexagonal lattice with hexagonal metallic patches. The side length of the patches is $R$, the side-to-side patches spacing is $g$, and the periodicity of the lattice in x and y directions is $D_x$ and $D_y$ respectively.

## APPENDIX D

### THE CAPACITANCE AND THE INDUCTANCE OF THE DESIGNED FSS UNIT CELL

In this section the capacitance of the two capacitive layers, and the inductance of the inductive layer in between is calculated and plotted in Fig. 21. The geometrical parameters of the FSS layers are presented in Table I. The capacitance and the inductance of the realized layers are shown in Fig. 21. It is seen that the obtained values are perfectly close to the desired values in the whole frequency bandwidth of simulation ($C = 78\ fF, L = 1.66\ nH$). A negligible deviation from the exact expected values is anticipated due to the dependency of $\varepsilon_{eff}$ and $\mu_{eff}$ as well as the intercoupling of the adjacent elements to the frequency of operation.

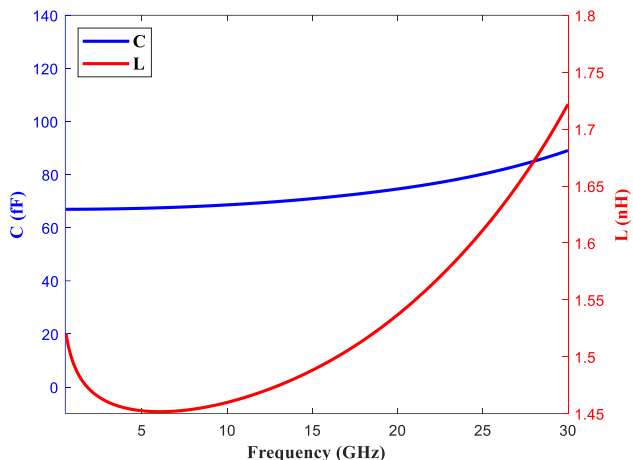

Fig. 21. The obtained values of the capacitance and the inductance of the layers of the proposed FSS depicted in Fig. 2. The left axis shows the capacitance of the blue curve, and the right axis shows the inductance of the red curve.

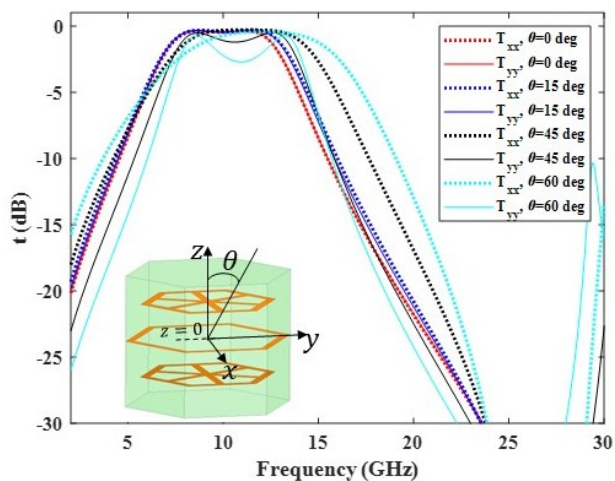

Fig. 22. Co-polarization transmission coefficient ($t = |S_{21}|$) of the FSS unit cell under different angles of incidence for x-polarized and y-polarized waves. The inset depicts the unit cell structure in which the angle of incidence $\theta$ is denoted.

## APPENDIX E

### THE EFFECT OF INCIDENCE ANGLE ON THE DESIGNED UNIT CELL RESPONSE

In this section the robustness of our designed unit cell to the incidence angle is investigated. A similar simulation setup to that of what we had for obtaining the unit cell response of the structure in section II, is assumed for this study. Figure 22 shows the co-polarization transmission coefficient $t = |S_{21}|$ spectrum for different incidence angles ($\theta = 0°, 15°, 45°, 60°$) for both x-polarized and y-polarized incident waves. The inset of Fig. 22 shows how the incidence angle $\theta$ is defined. As one can see, for normal incidence and small incidence angles, the structure response is almost the same for both polarizations. However, by increasing the incidence angle, the difference between the system response to the polarizations become more apparent and the system filtering performance is no longer polarization insensitive. It is also evident that by increasing the angle of incidence, the cut-off frequencies of the filter move towards higher frequencies leading to a shift in the filter operational bandwidth. This increase in the cut-off frequency is more noticeable for the higher cut-off frequency, and it happens more dominantly for the x-polarized waves. Nonetheless, the filter performance is still acceptable for large incidence angles as high as 45°.

## APPENDIX F

### DETERMINING THE VALUE OF THE $q$-PARAMETER OF THE RAISED-COSINE FEED MODEL

As explained in section III, we want to determine the values of $q$ in such a way so that the directivity of the feed model and its 3-dB beamwidth are matched to those of RF Spin QRH50 quad-ridged horn antennas used in the measurements. To do so, the parameter $q$ can be obtained by averaging the given values of directivity-match and beamwidth-match relations described respectively with [32]:

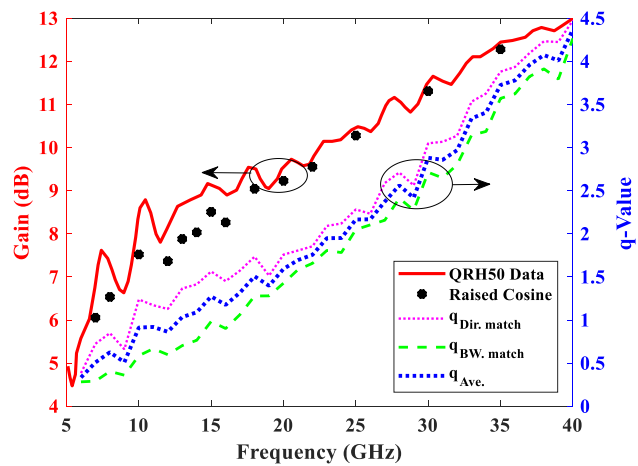

Fig. 23. The gain and the q-parameter values of the QRH50 horn antenna used in the measurements along with those of the raised-cosine feed model. The left axis depicts the gain values while the right axis depicts the q-parameter values.

$$q_{Dir.\,match}(f) = \frac{D_{QRH50}(f)}{4} - 0.5 \qquad (A8)$$

$$q_{BW\,match}(f) = \frac{-0.15}{\ln(\cos(\theta_{BW\,QRH50}(f)))} \qquad (A9)$$

where $D_{QRH50E}(f)$ and $\theta_{BW.QRH50E}(f)$ are the gain and 3-dB beamwidth spectra of the RF Spin QRH50 horns respectively, taken from their datasheet. The obtained values of $q$ for either equation and the averaged value, along with the gain values of the actual horn antenna and the raised cosine model is plotted in Fig. 23. It is seen that the gain of the raised cosine model follows the gain values of the actual horn reasonably.

APPENDIX G

CONVERGENCE STUDY: FINDING THE MINIMUM REQUIRED SIMULATION ACCURACY TO REDUCE THE SIMULATION COSTS

In this section, we present the results of 3D simulation of the structure at the sample frequency of 8 GHz for different accuracies of simulations. CST Microwave Studio allows for setting the number of triangular meshes discretizing the structure for the simulation. The higher the number of mesh cells per wavelength, the higher the number of total mesh cells and hence the higher precision of the results. We performed the simulation with 1, 2, 5, 10, 15, 20, 25, 30, 35 and 40 mesh cells per wavelength and extracted the total number of mesh cells, absolute electric field value, and the normalized transmission coefficient at 8 GHz. The total number of mesh cells on the hemispherical FSS for different values of mesh cells per wavelength is shown in Fig 24.

This increase translates to both higher accuracy of the results and simulation time. Obviously, to avoid time-costing studies, one ought to determine the minimum required accuracy especially for the simulations at higher frequencies. The normalized transmission of the structure and absolute electric field values at 8 GHz are obtained and plotted in Fig. 25. It is observable that the total number of mesh cells increases by increasing the number of mesh cells per wavelength.

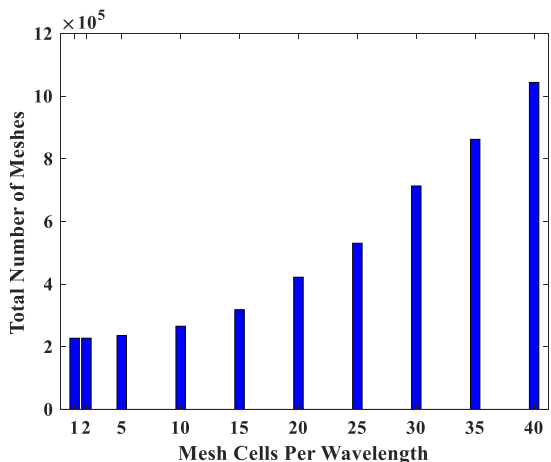
Fig. 24. Total number of meshes generated on the hemispherical structure in the IE-solver simulation process versus the numbers of the meshes per wavelength of simulation.

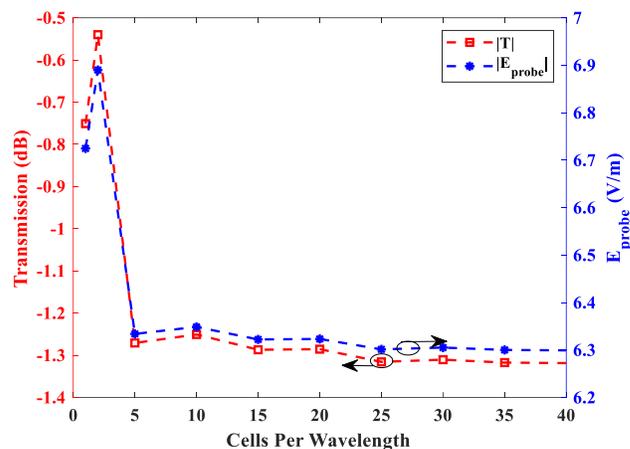
Fig. 25. Normalized transmission of the hemispherical structure in the boresight direction at 8 GHz, versus the numbers of the meshes per wavelength of simulation.

As expected, convergent behavior is seen with increasing the number of mesh cells per wavelength. However, we desire to lower the accuracy of the simulations as much as possible to reduce the simulation costs. If we choose 20 cells per wavelength, the relative error of the field value and the transmission coefficient, compared to the converged values, are less than 1 percent. This accuracy is a suitable tradeoff between the simulation costs and precision.

We also extended our convergence study to the structure gain pattern. The polar gain pattern of the hemispherical FSS at 8 GHz for 1, 15, 20 and 40 cells per wavelength in $x = 0$ plane ($\phi = 90°$) is plotted in Fig. 26. As anticipated, the gain pattern is also acceptably converged to the ultimate pattern with only 20 mesh cells per wavelength. Therefore, we conducted our numerical studies with 20 cells per wavelength in all our simulations.

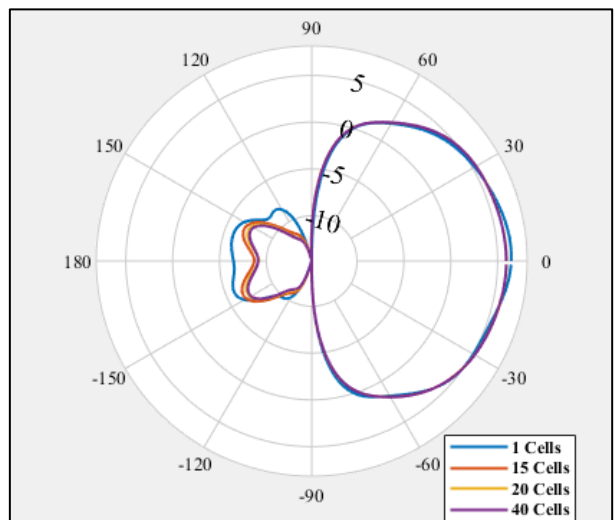
Fig. 26. Comparison of the polar gain patterns of the hemispherical structure at 8 GHz, in the $x = 0$ plane ($\phi = 90°$) for different simulation accuracies of 1, 15, 20 and 40 mesh cells per wavelength of simulation.

## APPENDIX H

### ABSOLUTE ELECTRIC FIELD SPECTRUM OF THE HEMISPHERICAL FSS IN THE BORESIGHT DIRECTION

Figure 27 depicts the obtained absolute electric field values at the location of the probe with the structure in place and without it. Figure 6 illustrates the simulation configuration for this study. One can see that as our design intended, the hemispherical FSS is allowing the passing of waves within the desired frequency range, although with some loss, and omits the transmission outside the passband region.

## APPENDIX I

### COMPARING THE 3D GAIN PATTERN OF THE HEMISPHERICAL FSS AT 10 GHz AND 20 GHz

The 3D gain patterns of the hemispherical structure at 10 GHz and 20 GHz are plotted in Figs. 28(a) and 28(b) respectively. Expectedly, we see a high transmission gain in the boresight direction at 10 GHz while a considerable reflection occurs towards the back direction at 20 GHz.

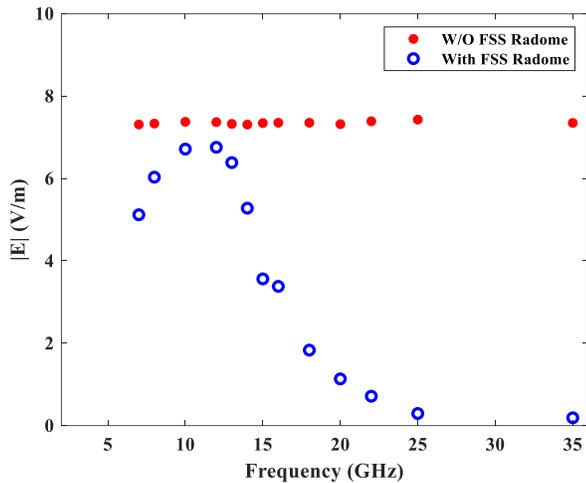

Fig. 27. The absolute electric field values at the probe location in the boresight direction versus frequency, for two scenarios once with hemispherical structure present, and once without the it in place.

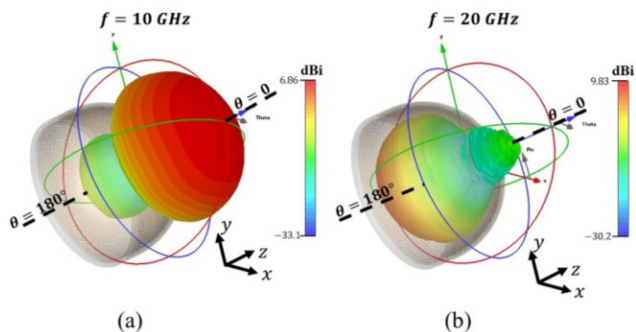

Fig. 28. 3D polar gain pattern of the hemispherical structure at 10 and 20 GHz. The reflective behavior of the structure at 20 GHz (in the band stop region) is obvious compared to the transmissive behavior at 10 GHz.